\documentclass[9pt,twocolumn,twoside]{opticajnl}

\journal{opticajournal} 

\setboolean{shortarticle}{true}

\title{Griffiths Anomalous Absorption in Sparse-Loss Photonic Lattices}

\author[1,2,*]{Stefano Longhi}

\affil[1]{Dipartimento di Fisica, Politecnico di Milano, Piazza L. da Vinci 32, I-20133 Milano, Italy}
\affil[2]{IFISC (UIB-CSIC), Instituto de Fisica Interdisciplinar y Sistemas Complejos - Palma de Mallorca, Spain}

\affil[*]{stefano.longhi@polimi.it}

\begin{abstract}
Light absorption in photonic lattices with sparsely distributed loss sites exhibits behavior analogous to Griffiths physics. Under uniform excitation, the transmitted power shows a stretched-exponential decay and a non-monotonic dependence on the loss strength, with an optimal loss rate that maximizes absorption. This behavior arises from rare, long loss-free segments that act as weakly coupled, long-lived photonic channels, rather than from exceptional point physics or interference effects.
Using a minimal tight-binding model with binary quenched dissipation, we show that rare regions produce a universal Griffiths-type subexponential decay. Sparse-loss photonic lattices thus provide an accessible platform to observe disorder-induced anomalous absorption and rare-region Griffiths physics.
\end{abstract}
\setboolean{displaycopyright}{false} 

\begin{document}

\maketitle

{\em Introduction.}  Controlling optical absorption is a longstanding objective in photonics~\cite{R1,R2,R3,R4}. A wide range of mechanisms have been developed to tailor absorption, from impedance matching and critical coupling in resonant structures~\cite{R5,R6,R7,R8,R9} to coherent perfect absorption and related interference-based schemes~\cite{R10,R11,R12,R13,R14,R15,R15b,R16,uffa}, as well as quantum-engineered approaches~\cite{Q1,Q2}. These methods rely on precise spatial or spectral design to enhance or suppress loss for applications in sensing, filtering, and signal processing. 
Recent advances in programmable loss landscapes, both in integrated photonic platforms and in fiber-loop lattices, have enabled controlled studies of non-Hermitian transport, Anderson localization, and light steering in extended dissipative systems~\cite{R18,R19,R20,R21,R22,R23,R24,R25,R26,R27,R28,R29,R30,R31,R32,R33,R34,R35}. Yet, the emergence of anomalous absorption in such settings has remained largely unexplored \cite{R18}.
Few-mode non-Hermitian structures can exhibit counterintuitive absorption effects, including non-monotonic loss dependence in two-mode optical "molecules" linked to exceptional points (EPs)~\cite{EP1,EP2,EP3}. {\color{black}However, in extended lattices with dissipation distributed over many sites, qualitatively different transport regimes emerge that are not captured by few-mode interference or EP physics.} In particular, lattices with sparse, randomly positioned absorbing sites~\cite{R31,R33} exhibit absorption behavior that stems from their disordered loss landscape, revealing mechanisms unique to extended non-Hermitian systems.

In this Letter, we show that quenched, spatially sparse dissipation leads to a striking stretched-exponential absorption and a non-monotonic dependence of transmitted power on the loss strength. {\color{black}While previous work~\cite{R33} predicted the formation of Lifshitz-tail photonic states, their consequences for optical absorption remained unexplored.} Here we identify the underlying mechanism as \emph{Griffiths rare-region physics}~\cite{G1,G2,G3,G4,G5,G6}, where exponentially rare loss-free segments dominate the long-distance dynamics. Griffiths phenomena are well known in disordered classical and quantum systems~\cite{G2,G4,G5,G6}, where they produce slow, non-analytic relaxation and broad lifetime distributions. Using a minimal tight-binding model with binary quenched loss, we analytically derive the resulting Griffiths-type stretched exponential absorption law {\color{black}and establish the direct connection between Lifshitz-tail states and Griffiths physics.} {\color{black}Moreover,} we demonstrate that the overall absorption is maximized at an optimal loss rate, {\color{black}a phenomenon that originates from rare-region statistics and is fundamentally distinct from the optimal absorption associated with EP physics.} Sparse-loss photonic lattices thus provide a simple, experimentally realistic platform where rare-region physics yields fundamentally non-exponential absorption in non-Hermitian optical systems.

{\em Model.}
We consider light propagation in a one-dimensional tight-binding waveguide lattice comprising $N$ coupled waveguides with a sparse landscape of optical loss \cite{R33}, schematically illustrated in Fig.~1(a). The same framework applies to other photonic platforms, such as synthetic temporal lattices \cite{R20,R25}.
Let $\gamma_n$ denote the loss rate at site $n$, and $J$ the coupling between adjacent waveguides. Light evolution of the light amplitudes $\psi_n(z)$ along the propagation coordinate $z$ is governed by the coupled-mode equations
\begin{equation}
i\frac{d\psi_n}{dz} = J(\psi_{n+1} + \psi_{n-1}) - i\,\gamma_n \psi_n,
\end{equation}
with open-boundary conditions $\psi_0(z)=\psi_{N+1}(z)=0$.
To model a sparse and disordered loss landscape, we assume that the $\gamma_n$ are independent quenched random variables drawn from a binary distribution, i.e. $\gamma_n=\gamma$ with probability $p$ and $\gamma_n=0$ with probability $(1-p)$.
For a given initial excitation of the lattice, $\psi_n(0)$ with normalization condition $\sum_n |\psi_n(0)|^2=1$, the fractional absorbed power after a propagation distance $z$ is given by $1-T(z,\gamma)$, where
\begin{equation}
T(z,\gamma)=\sum_n |\psi_n(z)|^2
\end{equation}
is the transmittance, which depends parametrically on the loss rate $\gamma$. Throughout this work, we typically consider uniform illumination at the input plane, i.e.,
$\psi_n(0)=1/\sqrt{N}$. The transmittance $T(z,\gamma)$ can be expressed as a sum of $2N$ exponentially decaying contributions via a spectral (supermode) decomposition of the coupled-mode equations, as detailed in Sec.~1 of the Supplemental document. For lattices with a relatively small number of waveguides, this multi-exponential structure generically leads to an asymptotic decay dominated by the least-damped supermode. In contrast, in the large-$N$ limit, a dense set of long-lived supermodes with comparable small damping rates emerges. In this regime, the simple single-mode dominance breaks down, and the resulting collective contribution leads to deviations from exponential decay and Griffiths physics.\\

 \begin{figure*}
 \centering
   \includegraphics[width=0.95\textwidth]{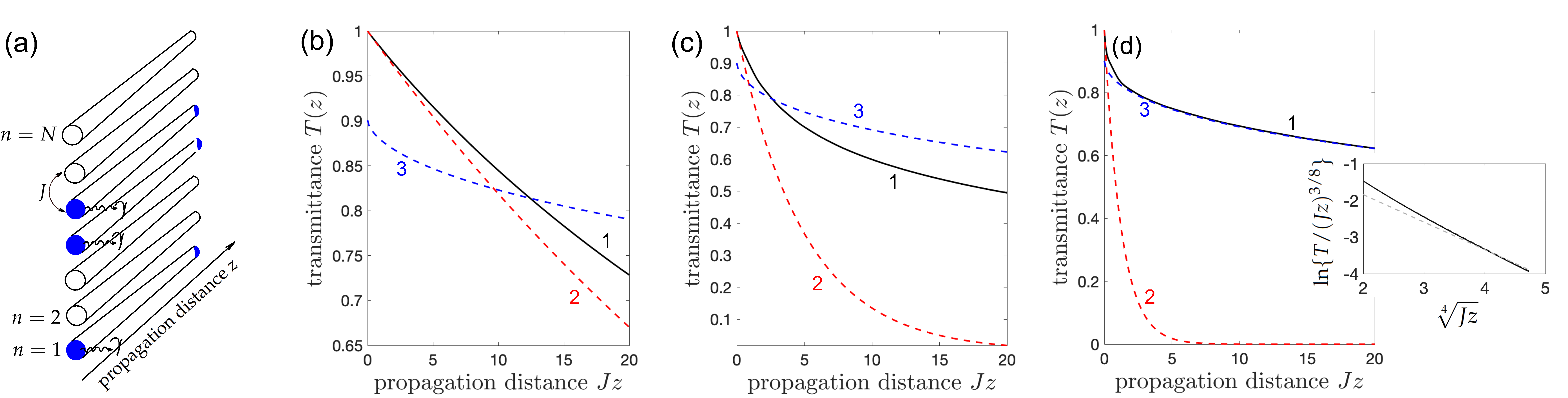}
   \caption{ \small (a) Schematic of a dissipative waveguide lattice comprising $N$ waveguides with sparse losses. Loss rate: $ \gamma$; coupling costant: $J$. (b-d) Numerically-computed behavior of the transmittance $T(z)$ versus normalized propagation distance $Jz$ (black curves 1) in a lattice comprising $N=50$ waveguides for increasing values of the normalized loss rate $\gamma / J$ and for $p=0.1$. (b) $\gamma/J=0.1$, (c) $\gamma/J=1$, and (d) $\gamma /J=4$. The curves are obtained after averaging over 1000 different realizations of disorder. The dashed red and blue curves (curves 2 and 3, respectively) show for comparison the exponential decay law predicted in the small lossy regime [Eq.(3)] and the stretched exponential decay in the strong loss regime [Eq.(4) with $\kappa=1.8$]. The inset in (d) shows the behavior of  $\ln \{  T / (Jz)^{3/8} \}$ versus $\sqrt[4]{Jz}$, clearly indicating the stretched exponential decay with exponent $\alpha=1/4$ as predicted by the asymptotic analysis. The dotted curve is the straight line with slope $C$ as predicted by Eq.(7).}
 \end{figure*}

 \begin{figure}
 \centering
   \includegraphics[width=0.5\textwidth]{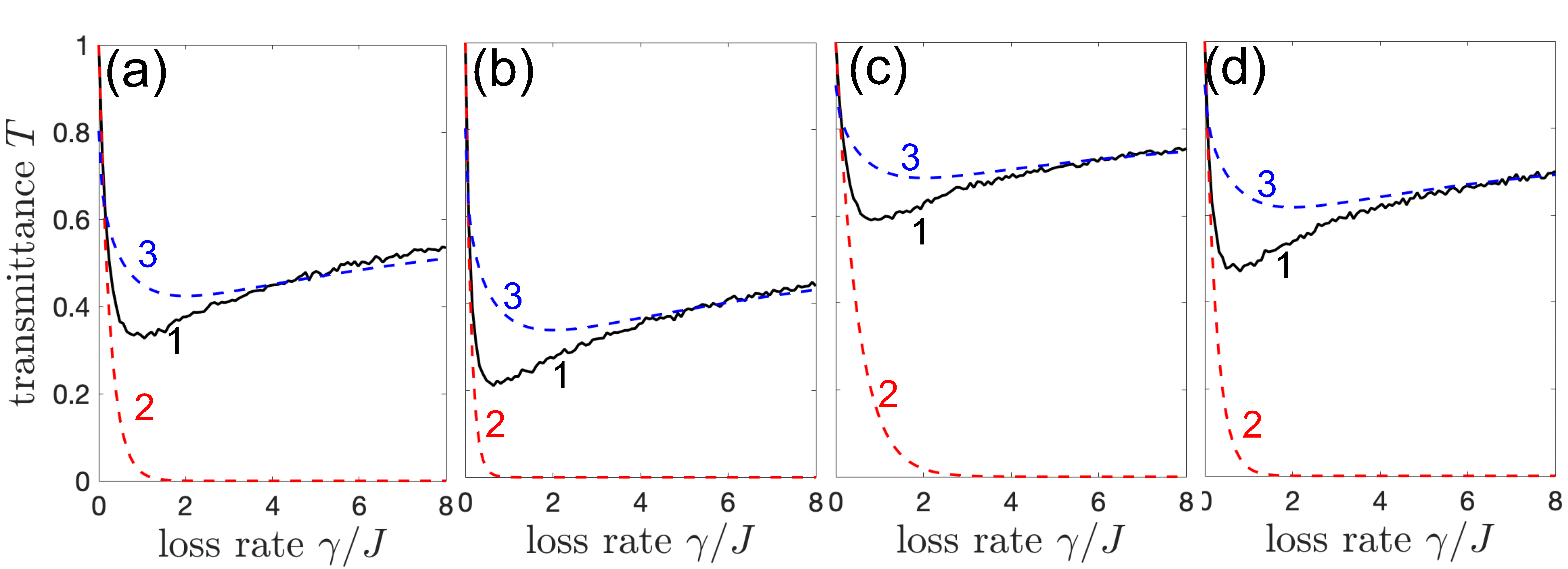}
   \caption{ \small Numerically-computed behavior of the transmittance $T(\gamma)$ versus normalized loss rate $ \gamma /J$ and fixed propagation distance $z$ (black curves 1) in a lattice comprising $N=50$ waveguides. Parameter values are: (a) $p=0.2$, $Jz=10$; (b) $p=0.2$, $Jz=20$; (c) $p=0.1$, $Jz=10$; (d) $p=0.1$, $Jz=20$.The dashed red and blue curves (curves 2 and 3, respectively) show for comparison the transmittance curves given by Eq.(3) (weak-loss limit) and Eq.(4) (strong-loss limit).}
 \end{figure}

{\em Anomalous Absorption from Griffiths rare regions.} 
When losses are uniformly distributed across the lattice ($p=1$), light undergoes standard Lambert--Beer attenuation, with transmittance 
$T(z,\gamma)=\exp(-2\gamma z)$.
A sparse and disordered loss landscape ($p\ll 1$) can fundamentally modify this behavior. As the loss strength increases from the weak to the moderate--strong regime, numerical simulations reveal a crossover from nearly exponential decay to a pronounced stretched--exponential attenuation. This trend is illustrated in Fig.~1(b--d), which displays the numerically computed transmittance $T(z,\gamma)$ versus propagation distance $z$ for several values of $\gamma/J$ in a lattice of size $N=50$ with $p=0.1$.
The qualitative origin of these distinct regimes is straightforward. For weak loss ($\gamma\ll J$), coupling between adjacent waveguides occurs over a scale $\sim 1/J$, much shorter than the absorption length $\sim 1/\gamma$. Light therefore spreads efficiently through the lattice before appreciable attenuation occurs, sampling lossy sites with a frequency set by their fraction $p$. In this limit, the dynamics is governed by an effective uniform absorption rate $\gamma_{\rm e}=p\gamma$, and the transmitted power follows approximately the Lambert--Beer law
\begin{equation}
T(z,\gamma) \simeq \exp(-2\gamma_e z);
\end{equation}
see Fig.2(b).
This simple picture, however, clearly breaks down in the moderate ($\gamma \sim J$) and strong ($\gamma \gg J$) dissipation regimes, as illustrated in Figs.1(c) and (d).{\color{black}For finite systems, the asymptotic decay at small $N$ is dominated by the lowest-decaying supermode, leading to an effectively single-exponential behavior, often accompanied by stopped absorption due to the emergence of dark states. Upon increasing $N$, however, a progressively denser set of weakly damped modes emerges, suppressing such finite-size saturation and enabling the onset of Griffiths-type physics, as shown in Figs.~S1 and S2 of the Supplemental Material.} In these regimes, absorption is governed by the structure and damping rates $\lambda$ of the supermodes of the disordered lattice.  
For $\gamma \gg J$, the spectrum of attenuation rates $\lambda$ separates into two groups~\cite{R33}: one with $\lambda \simeq \gamma$ (strongly attenuated modes) and another with $\lambda \sim J^2/\gamma$ (weakly attenuated modes).  At short propagation distances, light decays rapidly due to strongly damped supermodes, followed by a much slower decay associated with weakly damped ones. The long-distance (asymptotic) attenuation is dictated by the supermodes with the smallest loss rates, i.e., those with $\lambda$ closest to zero.  
These modes correspond to Lifshitz-tail states~\cite{R33}, which are the optical analogue of the rare-region excitations underlying Griffiths physics.  
They arise from exponentially rare, extended loss-free segments embedded in the dissipative lattice and form a hierarchy of increasingly long-lived, weakly attenuated photonic modes.  
In this sense, Lifshitz-tail states provide the concrete microscopic realization of the Griffiths mechanism: rare spatial regions dominate the long-time dynamics.  
In the large-$N$ limit, the asymptotic absorption is therefore controlled by the density of states and absorption rates of these rare, weakly damped supermodes, leading to a breakdown of the Lambert--Beer law and to subexponential decay; see Fig.1(c) and (d).

To gain analytical insight into this anomalous absorption and the breakdown of exponential decay in the moderate-to-strong loss regimes, we adopt a coarse-grained description of the lattice in terms of ``loss-free domains''---contiguous intervals free of dissipative sites. Each loss-free domain of length $\ell$ supports a set of long-lived localized photonic modes, closely analogous to the bound states of a finite potential box with approximately sinusoidal profiles \cite{R33}. Their leakage is governed by tunneling through the lossy sites at the domain boundaries.
The transmitted optical power $T(z,\gamma)$ can then be understood as arising from propagation through a sequence of statistically distributed loss-free domains embedded in a sparse absorbing lattice. After a short transient of order $1/\gamma$, all modes overlapping with lossy sites are rapidly attenuated, and the dynamics is dominated by light trapped in loss-free segments of typical length $\ell$. Each such domain contributes a quasi-bound mode whose decay rate $\lambda_\ell$ decreases rapidly with increasing $\ell$ due to the weak coupling to the surrounding lossy regions.
The total transmitted power is obtained by summing the contributions of all domains, each weighted by its length and statistical occurrence. {\color{black}In the large-$N$ limit, this sum can be replaced by a continuum average over domain lengths, leading to an effective integral over rare-region contributions governed by the Lifshitz-type statistics of long loss-free segments.} The resulting expression for the transmittance takes the form (technical details are given in Sec.~2 of the Supplemental document)
\begin{equation}
T(z,\gamma)\simeq
p^2 \int_0^\infty d\ell\;
\ell\,
\exp\!\left(
-\frac{\ell}{\xi}
-\frac{A J z}{\ell^3}
\right),
\end{equation}
where
\begin{equation}
\xi = -\frac{1}{\ln(1-p)},
\end{equation}
is the characteristic domain length, and
\begin{equation}
A = 
2\kappa \frac{4\pi^2 \gamma J}{\gamma^2+4J^2},
\end{equation}
encodes the effective boundary leakage rate, with $\kappa \gtrsim 1$ a dimensionless prefactor accounting for higher-order mode contributions.
For long propagation distances, the integral in Eq.(4) is evaluated by a saddle-point approximation, valid when the integrand becomes sharply peaked at large $z$. {\color{black}The saddle point corresponds to an optimal rare domain size resulting from the competition between the exponential suppression of long loss-free regions and their rapidly decreasing leakage rate.} This yields $\ell_* = (3A \xi J z)^{1/4}$ and therefore the asymptotic Griffiths-type behavior (see Sec.~2 of Supplemental document)
\begin{equation}
T(z,\gamma)\sim (Jz)^{3/8}
\exp\!\left(-C\, \sqrt[4] {Jz} \right),
\end{equation}
with $C \equiv (4/3^{3/4}) \times (A^{1/4}/ \xi^{3/4})$, showing that transport is governed by rare, long loss-free regions whose exponentially small leakage rates control the long-distance decay. {\color{black}The stretched-exponential exponent $\alpha=1/4$ arises in the present one-dimensional model from the interplay between the exponential statistics of rare domains (Lifshitz tail of domain lengths) and the algebraic decay of their leakage rates $\lambda_\ell \sim \ell^{-3}$. We note that different scaling exponents may arise in the presence of correlated disorder, long-range hopping, or higher-dimensional lattices, which alter either the domain statistics or the leakage law.} The numerical results shown in the inset of Fig.~1(d) demonstrate good agreement between the full simulations of $T(z,\gamma)$ and the asymptotic stretched-exponential behavior predicted by Eqs.(4) and (7).

{\em Optimal dissipation rate.} 
For a fixed propagation distance $z$, the total transmittance $T(z,\gamma)$ displays a clear minimum as a function of the loss parameter $\gamma$, thus revealing an optimal dissipation rate. This behavior can be understood by recalling that each quasi-bound mode with decay constant $\lambda_\ell(\gamma)$ contributes a factor $\exp[-\lambda_\ell(\gamma) z]$ to the total transmission. Since $\lambda_\ell(\gamma)$ increases linearly for small $\gamma/J$ but decreases as $1/(\gamma J)$ for large $\gamma/J$ [see Eq.(S9) in the Supplemental document], it necessarily reaches a maximum at an intermediate loss strength on the order of the coupling constant $J$. As a result, the overall transmission $T(z,\gamma)$ exhibits a corresponding minimum near $\gamma \sim J$, where absorption is most efficient. Physically, for weak dissipation the leakage is limited by the small imaginary potential, while for strong dissipation the excitation is rapidly repelled from the lossy boundaries, suppressing the coupling to the absorbing region and leading to loss-induced transparency. 
{\color{black}Unlike non-Hermitian few-mode systems, where non-monotonic absorption is typically associated with exceptional-point (EP) physics and spectral degeneracies ~\cite{EP1,EP2,EP3}, the present effect arises in an extended disordered lattice with no mode coalescence or EP structure. In large-dimensional or extended systems, EPs require highly fine-tuned parameter conditions and therefore represent measure-zero points in parameter space. In the sparse, quenched-disorder landscape considered here, such fine tuning is absent, making EP physics statistically irrelevant. The observed optimal absorption is instead a robust statistical effect arising from disorder-induced rare-region physics.}
Figure~2 illustrates the resulting non-monotonic dependence of $T$ on $\gamma$ for a few values of $p$ and output distance $z$, showing in each case a well-defined optimal loss strength.

 \begin{figure}
 \centering
   \includegraphics[width=0.45\textwidth]{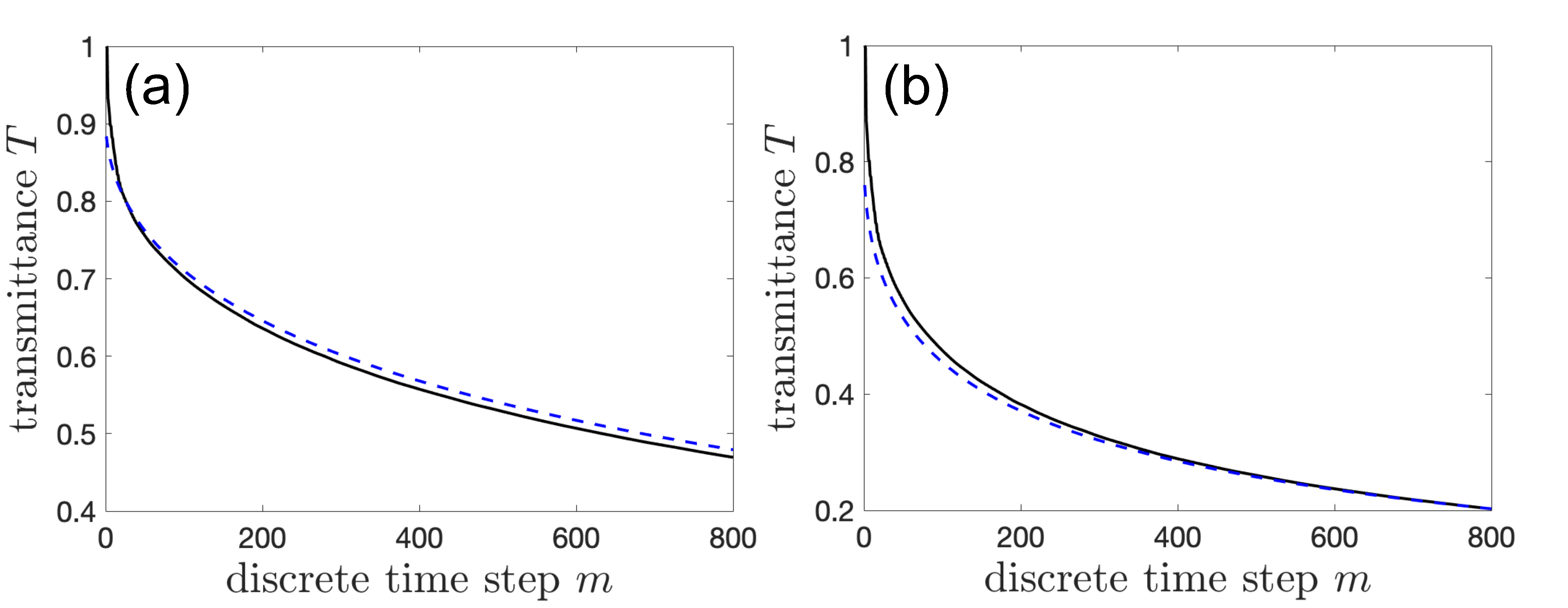}
   \caption{ \small Numerically-computed behavior of the transmittance $T$ versus propagation steps $m$ (solid curves) in a synthetic fiber loop lattice for $N=100$, $\beta= 0.9 \times \pi/2$, $\gamma=0.3$ and (a) $p=0.1$, (b) $p=0.2$. The curves are obtained after averaging over 1000 disorder realizations. The dashed curves show the theoretical predictions based on Eq.(4) with $J=(1/2) \cos \beta$.}
 \end{figure}

 {\it Anomalous absorption in synthetic temporal photonic lattices}. 
Synthetic temporal lattices realized in coupled fiber loops (see e.g. \cite{R18,R25,fiber,fiber2}) can provide an experimentally accessible platform for the observation of Griffiths-type anomalous absorption. The light dynamics of optical pulses circulating in the two fiber loops is governed by the discrete-time coupled equations \cite{R18,R24,R25,R33,fiber,fiber2}
\begin{eqnarray}
u_n^{(m+1)} & = & \left( \cos\beta\, u_{n+1}^{(m)} + i\sin\beta\, w_n^{(m)} \right)\exp(-2\gamma_n), \\[2mm]
w_n^{(m+1)} & = & \left( i\sin\beta\, u_n^{(m)} + \cos\beta\, w_{n-1}^{(m)} \right),
\end{eqnarray}
where $m$ is the round-trip number (the discrete analogue of the propagation distance $z$), $n$ labels the lattice site (pulse time slot), and $u_n^{(m)}$, $w_n^{(m)}$ denote the pulse amplitudes in the two loops. The site-dependent losses $\gamma_n$ are imposed in one loop via an amplitude modulator \cite{R24,fiber2} and are set to display a stochastic binary distribution. A finite lattice of $N$ unit cells is obtained by setting a coupling angle $\beta=\beta_0\neq\pi/2$ in the bulk and $\beta=\pi/2$ at the two edges. As in the spatial waveguide array, the long-time dynamics is controlled by Lifshitz modes with  longe lifetimes.
For a coupling angle $\beta$ close to $\pi/2$ and small $\gamma$, the synthetic temporal lattice effectively reproduces the tight-binding model (1), with hopping rate $J=\pm(1/2)\cos\beta$ and propagation distance identified with the round-trip number $z=m$ \cite{R31,fiber}. The transmittance is defined as $T(m)=\sum_n \left(|u_n^{(m)}|^2 + |w_n^{(m)}|^2\right)$, with $T(0)=1$. As an illustrative example, Fig.~3 shows the numerically computed behavior of $T$ versus $m$ for a fixed loss rate $\gamma$ in the strong-dissipation regime $\gamma\gg J$ and for two values of $p$. The synthetic lattice consists of $N=100$ unit cells, and we assume a uniform initial pulse excitation in one loop, i.e., $u_n^{(0)}=1/\sqrt{N}$ and $w_n^{(0)}=0$. Each transmittance curve is obtained after averaging over 1000 realizations of the quenched disorder in $\gamma_n$. The maximum round-trip number is $m=800$. These parameter values, including the ability to perform ensemble averaging, are fully within reach of present fiber-loop experiments \cite{R25,fiber2}.
The dashed curves in Fig.~3, obtained from the theoretical prediction [Eq.~(4) with $J=(1/2)\cos\beta$], show very good agreement with the numerical results, confirming that the fiber-loop platform provides an experimentally accessible setting for observing anomalous absorption and Griffiths-type physics.

{\em Conclusion.}  Photonic lattices with sparse quenched dissipation exhibit anomalous absorption governed by Griffiths rare-region physics, yielding stretched-exponential attenuation and a non-monotonic dependence of the transmittance on the loss strength. {\color{black} These effects originate solely from the statistics of the disordered loss landscape and are expected to persist in higher-dimensional lattices. Beyond revealing a new manifestation of Griffiths physics in optics, our results suggest a novel strategy for engineering optical absorption through the spatial statistics of loss, rather than through coherent interference or exceptional-point tuning. This disorder-assisted approach may provide new opportunities for optimizing integrated optical absorbers, tailoring dissipative photonic devices, and enhancing the sensitivity of loss-based optical sensing schemes.}\\

\noindent
{\bf Disclosures}. The author declares no conflicts of interest.\\
{\bf Data availability}. No data were generated or analyzed in the presented research.\\
{\bf Funding}. Agencia Estatal de Investigacion (MDM-2017-0711).\\
{\bf Supplemental document}. See Supplement 1 for supporting content.

\newpage


 {\bf References with full titles}\\
 \\
 \noindent
1. C.M. Watts, X. Liu, and W.J. Padilla,
Metamaterial Electromagnetic Wave Absorbers, Adv. Mater. {\bf 24}, OP98 (2012).\\
2. P. Yu, L.V. Besteiro, Y. Huang, J. Wu, L. Fu, H.H. Tan, C. Jagadish, G.P. Wiederrecht, A.O. Govorov, and Z. Wang, Broadband Metamaterial Absorbers, 
Adv. Opt. Mat. {\bf 7}, 1800995 (2019).\\
3. Q. Li, J. Lu, P. Gupta, and M. Qiu,
Engineering Optical Absorption in Graphene and Other 2D Materials: Advances and Applications, Adv. Opt. Mat. {\bf 7}, 1900595 (2019).\\
4. C. Ng, L. Wesemann, E. Panchenko, J. Song, T.J. Davis, A. Roberts, and D.E. G\'omez, Plasmonic Near-Complete Optical Absorption and Its Applications, Adv. Opt. Mat. {\bf 7}, 1801660 (2019).\\
5. A. Yariv, Universal relations for coupling of optical power between microresonators and dielectric waveguides, Electron. Lett. {\bf 36}, 321 (2000).\\
6. M. Cai, O. Painter, and K.J. Vahala, Observation of Critical Coupling in a Fiber Taper to a Silica-Microsphere Whispering-Gallery Mode System,
Phys. Rev. Lett. {\bf 85}, 74 (2000).\\
7. T.J. Kippenberg, S.M. Spillane, and K.J. Vahala, Modal coupling in traveling-wave resonators, Opt. Lett. {\bf 27}, 1669 (2002).\\
8. J.R. Piper, V. Liu, and S. Fan,
Total absorption by degenerate critical coupling,
Appl. Phys. Lett. {\bf 104}, 251110 (2014).\\
9. J.R. Piper and S. Fan,
Total Absorption in a Graphene Monolayer in the Optical Regime by Critical Coupling with a Photonic Crystal Guided Resonance, ACS Photon. {\bf 1}, 347 (2014).\\ 
10. Y.D. Chong, L. Ge, H. Cao, and A.D. Stone,
Coherent perfect absorbers: time-reversed lasers.
Phys. Rev. Lett. {\bf 105}, 053901 (2010).\\
11. S. Longhi, PT-symmetric laser-absorber, Phys. Rev. A {\bf 82},
031801 (2010).\\
12. W. Wan, Y. Chong, L. Ge, H. Noh, A.D. Stone, and H. Cao, Time-reversed lasing and interferometric control of absorption,
Science {\bf 331}, 889 (2011).\\
13. H. Noh, Y. Chong, A.D. Stone, and H. Cao,
Perfect coupling of light to surface plasmons by coherent absorption,
Phys. Rev. Lett. {\bf 108}, 186805 (2012).\\
14. Y. Sun, W. Tan, H.-q. Li, J. Li, and H. Chen, 
Experimental Demonstration of a Coherent Perfect Absorber with PT Phase Transition,
Phys. Rev. Lett. {\bf 112}, 143903 (2014).\\
15. S. Longhi and L. Feng, 
PT-symmetric microring laser-absorber, Opt. Lett. {\bf 39}, 5026 (2014).\\
16. Z.J. Wong, Y.L. Xu, J. Kim, K. O'Brien, Y. Wang, L. Feng, and X. Zhang,
Lasing and anti-lasing in a single cavity,
Nature Photon. {\bf 10}, 796 (2016).\\
17. H. H\"orner, L. Wild, Y. Slobodkin, G. Weinberg, O. Katz, and S. Rotter,
Coherent Perfect Absorption of Arbitrary Wavefronts at an Exceptional Point,
Phys. Rev. Lett. {\bf 133}, 173801 (2024).\\
18.  S. Longhi,  Algebraic Absorption in Non-Hermitian Photonic Lattices,
Photonics {\bf 13}, 574 (2026).\\
19. K.D.B. Higgins, S.C. Benjamin, T.M. Stace, G.J. Milburn, B.W. Lovett, and E.M. Gauger, 
Superabsorption of light via quantum engineering, 
Nature Commun. {\bf 5}, 4705 (2014).\\
20. D. Yang, S.-h. Oh, J. Han, G. Son, J. Kim, J. Kim, M. Lee, and K. An, Realization of superabsorption by time reversal of superradiance, Nature Photon. {\bf 15}, 272-276 (2021).\\
21. A. Regensburger,C. Bersch, B. Hinrichs, G. Onishchukov, A. Schreiber,
C. Silberhorn, and U. Peschel, 
Photon Propagation in a Discrete Fiber Network: An Interplay of Coherence and Losses, Phys. Rev. Lett. {\bf 107}, 233902 (2011).\\
22. A. Basiri , Y. Bromberg , A. Yamilov , H. Cao , and T. Kottos, Light localization induced by a random imaginary refractive index, Phys. Rev. A {\bf 90}, 043815 (2014).\\
23. S. Longhi, Topological Phase Transition in non-Hermitian Quasicrystals, Phys. Rev. Lett. {\bf 122}, 237601 (2019).\\
24. K.G. Makris, I. Kresic, A. Brandst\"otter, and S. Rotter, Scattering-free channels of invisibility across non-Hermitian media, Optica {\bf 7}, 619 (2020).\\
25. A. F. Tzortzakakis, K.G. Makris, and E.N. Economou, Non-Hermitian disorder in two-dimensional optical lattices, Phys. Rev. B {\bf 101}, 014202 (2020).\\
26. A. F. Tzortzakakis , K.G. Makris, A. Szameit, and E.N. Economou ,Transport and spectral features in non-Hermitian open systems, Phys. Rev. Res. {\bf 3}, 013208 (2021).\\
27. X. Luo, T. Ohtsuki, and R. Shindou, Universality Classes of the Anderson Transitions Driven by Non-Hermitian Disorder, Phys Rev. Lett. {\bf 126}, 090402 (2021). \\
28. S. Weidemann , M. Kremer , S. Longhi , and A. Szameit, Coexistence of dynamical delocalization and spectral localization through stochastic dissipation, Nat. Photonics {\bf 15}, 576 (2021).\\
29. H. Sahoo , R. Vijay , and S. Mujumdar, Anomalous transport regime in a non-Hermitian Anderson-localized hybrid system, Phys. Rev. Res. {\bf 4}, 043081 (2022).\\
30. A. Leventis, K.G. Makris, and E.N. Economou,  Non-Hermitian jumps in disordered lattices,
Phys. Rev. B {\bf 106}, 064205 (2022).\\
31. S. Longhi, Non-Hermitian control of localization in mosaic photonic lattices, App. Phys. Lett. {\bf 123}, 161102 (2023).\\ 
32. C. Wang and X.R. Wang, Anderson localization transitions in disordered non-Hermitian systems with exceptional points, Phys. Rev. B {\bf 107}, 024202 (2023).\\
33. S. Longhi, Anderson Localization in Dissipative Lattices, Ann. Phys. (Berlin) {\bf 535}, 2200658 (2023).\\
34. S. Longhi, Photonic random walks with traps, Opt. Lett. {\bf 49}, 2809 (2024).\\
35. B. Li,  C. Chen, and Z. Wang, Universal non-Hermitian transport in disordered systems, Phys. Rev. Lett. {\bf 135}, 033802 (2025).\\
36. S. Longhi, Lifshitz tail states in non-Hermitian disordered photonic lattices, Opt. Lett. {\bf 50}, 746 (2025).\\
37.  H. Zhao, X. Qiao, T. Wu, B. Midya, S. Longhi, and L. Feng,
Non-Hermitian topological light steering,
Science {\bf 365}, 1163 (2019).\\
38. Y. Feng, Z. Liu1, F. Liu, J. Yu, S. Liang, F. Li, Y. Zhang, M. Xiao, and Z. Zhang,
Loss Difference Induced Localization in a Non-Hermitian Honeycomb Photonic Lattice, Phys. Rev. Lett. {\bf 131}, 013802 (2023).\\
39. A. Guo, G. J. Salamo, D. Duchesne, R. Morandotti, M. Volatier-Ravat, V. Aimez, G. A. Siviloglou, and D. N. Christodoulides, Observation of PT-symmetry breaking in complex optical potentials, Phys. Rev. Lett. {\bf 103}, 093902 (2009).\\
40. B. Peng, K. \"{}Ozdemir, S. Rotter, H. Yilmaz, M. Liertzer, F. Monifi, C. M. Bender, F. Nori, and L. Yang, Loss-induced suppression and revival of lasing, Science {\bf 346}, 328 (2014).\\
41. I. Beder  and P. A. Brand\~ao, Quantum theory of loss-induced transparency in coupled waveguides, Phys. Rev. A {\bf 110}, 033503 (2024).\\
42. R. B. Griffiths, Nonanalytic behavior above the critical point in a random Ising ferromagnet, Phys. Rev. Lett. {\bf 23}, 17 (1969).\\
43. T. Vojta, Rare region effects at classical, quantum and nonequilibrium phase transitions, J. Phys. A {\bf 39}, R143 (2006).\\
44. M.A. Mu\~{n}oz, R. Juh\'asz, C. Castellano, and G. \'{O}dor,
Griffiths Phases on Complex Networks, Phys. Rev. Lett. {\bf 105}, 128701 (2010).\\
45. S. Gopalakrishnan, K. Agarwal, E. A. Demler, D. A. Huse, and M. Knap, Griffiths effects and slow dynamics in nearly many-body localized systems, Phys. Rev. B {\bf 93}, 134206 (2016).\\
46. K. Agarwal, E. Altman, E. A. Demler, S. Gopalakrishnan, D. A. Huse, and M. Knap, Rare-region effects and dynamics near the many-body localization transition, Ann. Phys. {\bf 529}, 1600326 (2017).\\
47. P.D. Bhoyar and P.M. Gade, Dynamic phase transition in the contact process with spatial disorder: Griffiths phase and complex persistence exponents, Phys. Rev. E {\bf 101}, 022128 (2020).\\
48. S. Wang, C. Qin, W. Liu, B. Wang, F. Zhou, H. Ye, L. Zhao, J. Dong, X. Zhang, S. Longhi, and P. Lu,
High-order dynamic localization and tunable temporal cloaking in ac-electric-field driven synthetic lattices,
Nature Commun. {\bf 13}, 7653 (2022).\\
49. L. Zhao, S. Wang, C. Qin, B. Wang, H. Ye, W. Liu, S. Longhi, and P. Lu, Real-time measurement of non-Hermitian
Landau-Zener tunneling near band crossings, Adv. Photon. {\bf 7}, 036002 (2025).


\begin{thebibliography}{99}


\bibitem{R1}
C.M. Watts, X. Liu, and W.J. Padilla, Adv. Mater. {\bf 24}, OP98 (2012).
\bibitem{R2}
P. Yu, L.V. Besteiro, Y. Huang, J. Wu, L. Fu, H.H. Tan, C. Jagadish, G.P. Wiederrecht, A.O. Govorov, and Z. Wang, Adv. Opt. Mat. {\bf 7}, 1800995 (2019).
\bibitem{R3}
Q. Li, J. Lu, P. Gupta, and M. Qiu,  Adv. Opt. Mat. {\bf 7}, 1900595 (2019).
\bibitem{R4}
C. Ng, L. Wesemann, E. Panchenko, J. Song, T.J. Davis, A. Roberts, and D.E. G\'omez,  Adv. Opt. Mat. {\bf 7}, 1801660 (2019).


\bibitem{R5}
A. Yariv, Electron. Lett. {\bf 36}, 321 (2000).
\bibitem{R6}
2. M. Cai, O. Painter, and K.J. Vahala, 
Phys. Rev. Lett. {\bf 85}, 74 (2000).
\bibitem{R7}
T.J. Kippenberg, S.M. Spillane, and K.J. Vahala, Opt. Lett. {\bf 27}, 1669 (2002).
\bibitem{R8}
J.R. Piper, V. Liu, and S. Fan, Appl. Phys. Lett. {\bf 104}, 251110 (2014).
\bibitem{R9}
J.R. Piper and S. Fan,  ACS Photon. {\bf 1}, 347 (2014). 



\bibitem{R10}
Y.D. Chong, L. Ge, H. Cao, and A.D. Stone,
Phys. Rev. Lett. {\bf 105}, 053901 (2010).
\bibitem{R11}
S. Longhi, Phys. Rev. A {\bf 82},
031801 (2010).
\bibitem{R12}
W. Wan, Y. Chong, L. Ge, H. Noh, A.D. Stone, and H. Cao, 
Science {\bf 331}, 889 (2011).
\bibitem{R13}
H. Noh, Y. Chong, A.D. Stone, and H. Cao,
Phys. Rev. Lett. {\bf 108}, 186805 (2012).
\bibitem{R14}
Y. Sun, W. Tan, H.-q. Li, J. Li, and H. Chen, 
Phys. Rev. Lett. {\bf 112}, 143903 (2014).
\bibitem{R15}
S. Longhi and L. Feng, Opt. Lett. {\bf 39}, 5026 (2014).
\bibitem{R15b}
Z.J. Wong, Y.L. Xu, J. Kim, K. O'Brien, Y. Wang, L. Feng, and X. Zhang,
Nature Photon. {\bf 10}, 796 (2016).
\bibitem{R16}
H. H\"orner, L. Wild, Y. Slobodkin, G. Weinberg, O. Katz, and S. Rotter,
Phys. Rev. Lett. {\bf 133}, 173801 (2024).

\bibitem{uffa}
S. Longhi,  Photonics {\bf 13}, 574 (2026).


\bibitem{Q1}
K.D.B. Higgins, S.C. Benjamin, T.M. Stace, G.J. Milburn, B.W. Lovett, and E.M. Gauger, Nature Commun. {\bf 5}, 4705 (2014).
\bibitem{Q2}
D. Yang, S.-h. Oh, J. Han, G. Son, J. Kim, J. Kim, M. Lee, and K. An, Nature Photon. {\bf 15}, 272-276 (2021).







\bibitem{R18}
A. Regensburger,C. Bersch, B. Hinrichs, G. Onishchukov, A. Schreiber,
C. Silberhorn, and U. Peschel, Phys. Rev. Lett. {\bf 107}, 233902 (2011).
\bibitem{R19}
A. Basiri , Y. Bromberg , A. Yamilov , H. Cao , and T. Kottos, Phys. Rev. A {\bf 90}, 043815 (2014).
\bibitem{R20}
S. Longhi, Phys. Rev. Lett. {\bf 122}, 237601 (2019).
\bibitem{R21}
K.G. Makris, I. Kresic, A. Brandst\"otter, and S. Rotter, Optica {\bf 7}, 619 (2020).
\bibitem{R22}
A. F. Tzortzakakis, K.G. Makris, and E.N. Economou, Phys. Rev. B {\bf 101}, 014202 (2020).
\bibitem{R23}
A. F. Tzortzakakis , K.G. Makris, A. Szameit, and E.N. Economou, Phys. Rev. Res. {\bf 3}, 013208 (2021).
\bibitem{R24}
X. Luo, T. Ohtsuki, and R. Shindou,  Phys Rev. Lett. {\bf 126}, 090402 (2021). 
\bibitem{R25}
S. Weidemann , M. Kremer , S. Longhi , and A. Szameit,  Nature Photon. {\bf 15}, 576 (2021).
\bibitem{R26}
H. Sahoo , R. Vijay , and S. Mujumdar,  Phys. Rev. Res. {\bf 4}, 043081 (2022).
\bibitem{R27}
A. Leventis, K.G. Makris, and E.N. Economou, Phys. Rev. B {\bf 106}, 064205 (2022).
\bibitem{R28}
S. Longhi,  App. Phys. Lett. {\bf 123}, 161102 (2023).
\bibitem{R29}
C. Wang and X.R. Wang, Phys. Rev. B {\bf 107}, 024202 (2023).
\bibitem{R30}
S. Longhi,  Ann. Phys. (Berlin) {\bf 535}, 2200658 (2023).
\bibitem{R31}
S. Longhi,  Opt. Lett. {\bf 49}, 2809 (2024).
\bibitem{R32}
B. Li,  C. Chen, and Z. Wang, Phys. Rev. Lett. {\bf 135}, 033802 (2025).
\bibitem{R33}
S. Longhi, Opt. Lett. {\bf 50}, 746 (2025).
\bibitem{R34}
H. Zhao, X. Qiao, T. Wu, B. Midya, S. Longhi, and L. Feng,
Science {\bf 365}, 1163 (2019).
\bibitem{R35}
Y. Feng, Z. Liu1, F. Liu, J. Yu, S. Liang, F. Li, Y. Zhang, M. Xiao, and Z. Zhang, Phys. Rev. Lett. {\bf 131}, 013802 (2023).

\bibitem{EP1}
A. Guo, G. J. Salamo, D. Duchesne, R. Morandotti, M. Volatier-Ravat, V. Aimez, G. A. Siviloglou, and D. N. Christodoulides, Phys. Rev. Lett. {\bf 103}, 093902 (2009).
\bibitem{EP2}
B. Peng, K. \"{O}zdemir, S. Rotter, H. Yilmaz, M. Liertzer, F. Monifi, C. M. Bender, F. Nori, and L. Yang, Science {\bf 346}, 328 (2014).
\bibitem{EP3}
I. Beder  and P. A. Brand\~ao,  Phys. Rev. A {\bf 110}, 033503 (2024).


\bibitem{G1}
R. B. Griffiths, Phys. Rev. Lett. {\bf 23}, 17 (1969).
\bibitem{G2}
T. Vojta, J. Phys. A {\bf 39}, R143 (2006).
\bibitem{G3}
M.A. Mu\~{n}oz, R. Juh\'asz, C. Castellano, and G. \'{O}dor, Phys. Rev. Lett. {\bf 105}, 128701 (2010).
\bibitem{G4}
S. Gopalakrishnan, K. Agarwal, E. A. Demler, D. A. Huse, and M. Knap, Phys. Rev. B {\bf 93}, 134206 (2016).
\bibitem{G5}
K. Agarwal, E. Altman, E. A. Demler, S. Gopalakrishnan, D. A. Huse, and M. Knap, Ann. Phys. {\bf 529}, 1600326 (2017).
\bibitem{G6}
P.D. Bhoyar and P.M. Gade, Phys. Rev. E {\bf 101}, 022128 (2020).

\bibitem{fiber}
S. Wang, C. Qin, W. Liu, B. Wang, F. Zhou, H. Ye, L. Zhao, J. Dong, X. Zhang, S. Longhi, and P. Lu, Nature Commun. {\bf 13}, 7653 (2022).
\bibitem{fiber2}
 L. Zhao, S. Wang, C. Qin, B. Wang, H. Ye, W. Liu, S. Longhi, and P. Lu, Adv. Photon. {\bf 7}, 036002 (2025).

\end{thebibliography}
\end{document}